\documentstyle[12pt]{article}
\begin{document}
%
\newcommand{\beqn}{\begin{equation}}
\newcommand{\eeqn}{\end{equation}}
\newcommand{\beqnar}{\begin{eqnarray}}
\newcommand{\eeqnar}{\end{eqnarray}}
\newcounter{abc}
\newcommand{\beqna}%
     {\renewcommand{\theequation}{\arabic{section}.\arabic{equation}\alph{abc}}
     \setcounter{abc}{1}
     \begin{equation}}
\newcommand{\eeqna}{\end{equation}
     \renewcommand{\theequation}{\arabic{section}.\arabic{equation}}}
\newcommand{\beqnb}%
     {\renewcommand{\theequation}{\arabic{section}.\arabic{equation}\alph{abc}}
     \setcounter{abc}{2}
     \addtocounter{equation}{-1}
     \begin{equation}}
\newcommand{\eeqnb}{\end{equation}
     \renewcommand{\theequation}{\arabic{section}.\arabic{equation}}}
\newcommand{\beqnc}%
     {\renewcommand{\theequation}{\arabic{section}.\arabic{equation}\alph{abc}}
     \setcounter{abc}{3}
     \addtocounter{equation}{-1}
     \begin{equation}}
\newcommand{\eeqnc}{\end{equation}
     \renewcommand{\theequation}{\arabic{section}.\arabic{equation}}}
\newcommand{\beqnd}%
     {\renewcommand{\theequation}{\arabic{section}.\arabic{equation}\alph{abc}}
     \setcounter{abc}{4}
     \addtocounter{equation}{-1}
     \begin{equation}}
\newcommand{\eeqnd}{\end{equation}
     \renewcommand{\theequation}{\arabic{section}.\arabic{equation}}}
\newcommand{\beqne}%
     {\renewcommand{\theequation}{\arabic{section}.\arabic{equation}\alph{abc}}
     \setcounter{abc}{5}
     \addtocounter{equation}{-1}
     \begin{equation}}
\newcommand{\eeqne}{\end{equation}
     \renewcommand{\theequation}{\arabic{section}.\arabic{equation}}}
\newcommand{\seta}{\setcounter{abc}{1}}
\newcommand{\setb}{\setcounter{abc}{2}\addtocounter{equation}{-1}}
\newcommand{\setc}{\setcounter{abc}{3}\addtocounter{equation}{-1}}
\newcommand{\setd}{\setcounter{abc}{4}\addtocounter{equation}{-1}}
\newcommand{\sete}{\setcounter{abc}{5}\addtocounter{equation}{-1}}
\newcommand{\beqnarlett}{
     \renewcommand{\theequation}{\arabic{section}.\arabic{equation}\alph{abc}}
     \begin{eqnarray}}
\newcommand{\eeqnarlett}{\end{eqnarray}
     \renewcommand{\theequation}{\arabic{section}.\arabic{equation}}}
%
%

\renewcommand{\theequation}{\arabic{section}.\arabic{equation}}
\renewcommand{\thefootnote}{\alph{footnote}}
\begin{flushright}
CERN-TH.6935/93
\end{flushright}
\begin{center}
{\bf \LARGE Finite Chains with Quantum Affine Symmetries}
\\*[1.5cm]
{\Large F. C. Alcaraz,\footnote{Departamento de F\'{\i}sica, Universidade
 Federal
de S\~{a}o Carlos, 13560 S\~{a}o Carlos-SP, Brazil}
\footnote{Physikalisches Institut, Bonn University, 53115 Bonn, Germany}
\hspace{0.5cm}D. Arnaudon,\footnote{Theory Division, CERN, 1211 Gen\`{e}ve 23,
Switzerland} \hspace{0.5cm} V. Rittenberg,$^b$ \\
M. Scheunert${^b}$}
\vspace{1cm}
\begin{abstract}
We consider an extension of the ($t$--$U$)
Hubbard model taking into account new
interactions between the numbers of up and down electrons. We confine ourselves
to a one-dimensional open chain with $L$ sites ($4^L$ states) and derive the
ef\/fective
Hamiltonian in the strong repulsion (large $U$) regime. This
Hamiltonian acts on
$3^L$ states. We show that the spectrum of the latter Hamiltonian (not the
degeneracies) coincides with the spectrum of the anisotropic Heisenberg chain
(XXZ model) in the presence of a $Z$ field ($2^L$ states).  The wave functions
of the $3^L$-state system are obtained explicitly from those of the $2^L$-state
system, and the degeneracies can be understood in terms of irreducible
representations of $U_q(\widehat{sl(2)})$.
\end{abstract}
\vspace*{1.cm}
\begin{flushleft}
CERN-TH.6935/93
\end{flushleft}
\end{center}
\newpage
\section{Introduction}
The one-dimensional Hubbard model, which depends on two
parameters ($U$ and $t$),
has received a lot of attention due to its possible application (in its 2-d
version) to describe high-$T_c$ oxide materials [1]. The large $U$ limit
of this system was discussed in Refs. [2].

In this paper we consider the following Hamiltonian:
\begin{equation}
H = H_0 + H_1 \; ,
\end{equation}
where
\beqnarlett
\seta
H_0  = & & \sum^{L-1}_{i=1}
\left\{- \sum_{\sigma = \uparrow , \downarrow}
\left( c^+_{i, \sigma}c_{i + 1, \sigma} - c_{i, \sigma} c^+_{i + 1, \sigma}
 \right) \right.
 + \frac{U}{2t} (n_i , _\uparrow n_{i , \downarrow} +
	n_{i + 1 , \uparrow} n_{i + 1 , \downarrow} ) \nonumber \\*[0.5cm]
& + &  \frac{V}{t} (n_i , _\uparrow - n_{i, \downarrow} )^2
	( n_{i + 1 , \uparrow} - n_{i + 1 , \downarrow})^2 \nonumber \\*[0.5cm]
& + &
\frac{W^\cdot}{2t} \left[(n_i , _\uparrow - n_{i, \downarrow} )^2
	+ ( n_{i + 1 , \uparrow} - n_{i + 1 , \downarrow})^2 \right] \nonumber
\\*[0.5cm]
& + &
\frac{A}{2t}  \left. \left[(n_i , _\uparrow - n_{i, \downarrow} )^2
	- ( n_{i + 1 , \uparrow} - n_{i + 1 , \downarrow})^2 \right]
+ \frac{B}{t} \right\}
\eeqnarlett
and
\beqnarlett
\setb
H_1 & = & \frac{G}{2t} \sum^L_{i = 1} ( n_{i , \uparrow} - n_{i , \downarrow})
 \; .
\eeqnarlett
Here $c^+_{i, \sigma} , c_{i, \sigma}$ and $n_{i, \sigma} = c^+_{i , \sigma}
c_{i, \sigma} (\sigma = \uparrow , \downarrow)$ are creation,
annihilation and occupation number fermionic operators. In
the Hubbard
model one has the hopping term (with coefficient $-1$),
the repulsion term (with coef\/ficient
$\frac{U}{2t}$) and the external field (1.2b) (coef\/ficient $\frac{G}{2t}$).
We have generalized this Hamiltonian by adding a quartic interaction term
(coef\/ficient $\frac{V}{t}$) and an external field type term (coef\/ficient
$\frac{W}{2t}$). The term proportional to $\frac{A}{2t}$ corresponds to a
boundary term; it does not change the thermodynamical properties of the system
 but,
as we shall see, plays an important role in its symmetry properties. For
similar reasons we have added the constant $\frac{B}{t}$, which will be shown
to be useful for the identification of representations of some Hecke algebras.
We have no phenomenological reasons to introduce
the supplementary terms in the Hamiltonian but, as will be seen, the phase
structure of the physical system thus obtained is much richer.

If, as in the $t$--$J$ model [1], we take the large $\frac{U}{t}$ limit
keeping $\frac{G}{2t} = g , \frac{V}{t} = v , \frac{W}{2t} = w ,
\frac{A}{2t} = a \; {\mbox{and}} \;  \frac{B}{t} = b$ fixed, as shown in
Appendix A, we obtain instead of a Hamiltonian acting on $4^L$ states, an
ef\/fective Hamiltonian acting on $3^L$ states since the remaining states get
an infinite energy and drop out of the spectrum. This ef\/fective Hamiltonian
reads
\beqn
H^\prime = H^\prime_0 + H^\prime_1 \; ,
\eeqn
where
\beqn
H^\prime_0  =  \sum^{L-1}_{i = 1} U_i
\eeqn
and
\begin{eqnarray}
U_i  = & - &
      \left( \varrho^-_i \varrho^+_{i + 1} + \varrho^+_i \varrho^-_{i + 1}
	+ \tau^+_i \tau^-_{i + 1} + \tau^-_i \tau^+_{i + 1} \right)
\nonumber \\*[0.5cm]
	& + & v \varepsilon^0_i \varepsilon^0_{i + 1} + w (\varepsilon^0_i +
	    \varepsilon^0_{i+1} ) \nonumber \\*[0.5cm]
	& + & a ( \varepsilon^0_i - \varepsilon^0_{i + 1}) + b
\end{eqnarray}
and
\beqn
H^\prime_1 = - g \sum^L_{i = 1} \varepsilon^z_i \; .
\eeqn
We use here the notation of Ref. [3]:
\[
\varrho^+ = \begin{array}{c}
\left( \begin{array}{ccc}
		0 & 0 & 1 \\
		0 & 0 & 0 \\
		0 & 0 & 0   \end{array} \right),\qquad
\varrho^- = \left( \begin{array}{ccc}
		0 & 0 & 0 \\
		0 & 0 & 0 \\
		1 & 0 & 0   \end{array} \right),\qquad
\tau^+ = \left( \begin{array}{ccc}
		0 & 1 & 0 \\
		0 & 0 & 0 \\
		0 & 0 & 0  \end{array} \right),
\end{array}
\]
\[
\tau^- = \begin{array}{c}
\left( \begin{array}{ccc}
		0 & 0 & 0 \\
		1 & 0 & 0 \\
		0 & 0 & 0  \end{array} \right),\qquad
\varepsilon^0 = \left( \begin{array}{ccc}
		0 & 0 & 0 \\
		0 & 1 & 0 \\
		0 & 0 & 1   \end{array} \right),\qquad
\varepsilon^z = \left( \begin{array}{ccc}
		0 & 0 & 0 \\
		0 & 1 & 0 \\
		0 & 0 & -1   \end{array} \right),
\end{array}
\]
\beqnar
\varepsilon^+ = \begin{array}{c}
\left( \begin{array}{ccc}
		0 & 0 & 0 \\
		0 & 0 & 1 \\
		0 & 0 & 0  \end{array} \right),\qquad

\varepsilon^- = \left( \begin{array}{ccc}
		0 & 0 & 0 \\
		0 & 0 & 0 \\
		0 & 1 & 0  \end{array} \right) \; .
\end{array}
\eeqnar
Obviously:
\beqn
[ H^\prime_0 , S^z ] = [ H^\prime_0 , S^+ ] = [ H^\prime_0 , S^- ] =
[ H^\prime_0 , S^0 ] = 0 \; ,
\eeqn
where
\beqn
S^z = \frac{1}{2} \sum^L_{i = 1} \varepsilon^z_i     \: ; \:
S^\pm = \sum^L_{i = 1} \varepsilon^\pm_i \: ; \:
S^0   = \sum^L_{i = 1} \varepsilon^0_i \; .
\eeqn

Thus $H^\prime_0$ is $SU(2) \otimes U(1)$ invariant (generators $S^\pm , S^z$
 and
$S^0$). As will be seen later, the symmetry algebra is much larger, namely
$H^\prime_0$ is invariant under $U_q (sl(2)) \otimes U(1)$ for any value of
 $q\:$!

Special cases of the quantum chain (1.4) were already obtained in a
 mathe\-matical
context. In Ref. [4] it was noticed that if
\beqn
v = 0 \: , \: w + a = q \: , \: w-a = q^{-1} \: , \: b = 0
\eeqn
the $U_j$'s are generators of the Hecke algebra. We notice that if we take:
\beqn
v = 0 \: , \: w = 1 \: , \: a = 0
\eeqn
we obtain a special case of the one-parameter-dependent $U_q(sl(2)) (q^3 = 1)$
invariant quantum chain of Ref. [3]. In this model one uses nilpotent
representations. The same quantum chain plays an important role in a physical
application [5].
If one considers the one-dimensional problem in which two types of molecules,
say A and B, dif\/fuse in a lattice (vacancies represent the third state of the
problem) and undergo chemical reactions, the time evolution operator
that
appears in the master equation is given, for certain rates, by the
Hamiltonian (1.4). We
stress that in order to obtain Eq. (1.4) the molecules A and B have to play a
symmetric role.

It is the aim of this paper to study the spectrum and
eigenfunctions of the Hamiltonian $H^\prime_0$. If this problem is solved,
since $H^\prime_1$ commutes with $H^\prime_0$, the spectrum
of $H^\prime$ is trivially obtained for finite chains.
The calculation in the thermodynamical
limit of the spectrum and correlation functions in the presence of $H^\prime_1$
has to be done, as usual, with care, since the ground state of the system is
dependent on the coupling constant $g$ appearing in Eq. (1.6). As will be shown
in the present paper, the spectrum and wave functions of the Hamiltonian
$H^\prime_0$ can be obtained from those of another Hamiltonian, namely the XXZ
 spin $\frac{1}{2}$ Heisenberg
chain in the presence of a $Z$ field, which is integrable. The last
Hamiltonian depends (leaving aside the boundary terms) on the anisotropy
$\Delta$ and the strength $h$ of the $Z$ field. These two parameters are
related to the two parameters $v$ and $w$ in Eq. (1.5). It turns out that the
energy levels of the Heisenberg chain ($2^{L}$ states) and those of the
chain given by Eq. (1.4) ($3^L$ states) are identical. The degeneracies
of the $3^L$ states chain depend on those of the $2^L$ chain, which in turn
depend on the values of $\Delta , h$ and boundary terms.

The paper is organized as follows. In Sec. 2 we shortly review the phase
diagram of the Heisenberg chain since the Hamiltonian (1.4) has the same phase
diagram. We also remind the reader of the known symmetries of the model
(quantum group symmetries) and point to an amusing new symmetry, which has
as consequence the degeneracy of just two special energy levels for any finite
length (see also Appendix B). We then rewrite the
Heisenberg model in a dif\/ferent basis and show how to get other
quantum chains
with the same spectrum, the Hamiltonian (1.4) being the simplest case.

In Sec. 3 we give a simple construction that allows, starting from an
eigenfunction of the Heisenberg chain corresponding to a given eigenvalue,
to get all the corresponding eigenfunctions of the $3^L$ states chain
with the same energy.

In Sec. 4 we look at this problem from a dif\/ferent point of view. Starting
from the observation that $H^\prime_0$ is invariant under the
$SU(2)$ given by the
generators $S^\pm$ and $S^z$, we try to $q$-deform the chain and look for a
Hamiltonian that commutes with the known generators $\tilde{S}^\pm$ and $S^z$
of the quantum algebra $U_q (sl(2))$ (for $q = 1 , \tilde{S}^\pm$
coincides with
$S^\pm$). Surprisingly enough, the chain stays $q$-independent although the
generators $\tilde{S}^\pm$ are (using the coproduct rules) $q$-dependent. This
implies that the chain is invariant under $U_q(sl(2))$ for any $q$, in
particular it is invariant under the af\/fine $U_q(\widehat{sl(2)})$ algebra
for any $q$. In $U_q\widehat{(sl(2))}$ one combines $U_q(sl(2))$
with $U_{q^{-1}}(sl(2))$. This allows us to understand the
degeneracies of the (1.4) chain in terms of finite-dimensional irreducible
representations of $U_q(\widehat{sl(2)})$.

Starting from this observation, in Sec. 5 (see also Appendix C) we ask under
 which circumstances a
$3^L$ chain can be invariant under $U_q(\widehat{sl(2)})$,
if we take a spin $0$ and a spin $\frac{1}{2}$ on each site.
The answer we get is
again the Hamiltonian (1.4). We warn the reader that, as opposed to the first
sections, the last two (4 and 5) are rather mathematical in
character.

Finally, in Sec. 6 we describe the degeneracies of the $3^L$ chain if the
underlying Heisenberg chain has supplementary symmetries as described in
Sec. 2 and Appendix B.
Our conclusions are presented in Sec. 7.

\section{A procedure to obtain quantum chains with the same spectrum as the
Heisenberg chain}
\setcounter{equation}{0}
The spin $\frac{1}{2}$
anisotropic Heisenberg chain is defined by the Hamiltonian
\begin{equation}
H^{\prime \prime} = \sum^{L-1}_{i=1} V_i \; ,
\end{equation}
where
\begin{equation}
V_i = - \frac{1}{2} \left[
\sigma^x_i \sigma^x_{i + 1} + \sigma^y_i  \sigma^y_{i + 1} +
\Delta \sigma^z_i \sigma^z_{i + 1} + h(\sigma^z_i + \sigma^z_{i + 1})
 + \alpha (\sigma^z_i - \sigma^z_{i + 1}) + \beta \right]
\end{equation}
and $\sigma^x , \sigma^y$ and $\sigma^z$ are Pauli matrices. As is
well known,
 this
chain is integrable with the help of the Bethe ansatz. If $\Delta, h$ and
$\alpha$ are arbitrary, there are no degeneracies in the spectrum. Obviously,
\begin{equation}
\left[ I^z , H^{\prime \prime} \right] = 0 \; ,
\end{equation}
where
\begin{equation}
I^z = \frac{1}{2} \sum^L_{i = 1} \sigma^z_i \; .
\end{equation}
One obtains higher symmetries in three cases: \\ \\
{\it Case 1a}
\begin{equation}
\Delta = 0 ,\;
h + \alpha = - q ,\;
h - \alpha = - q^{-1} ,\;
\beta = - (q - q^{-1}) \;.
\end{equation}
In this case the Hamiltonian is $U_q(sl(1/1))$-symmetric [7] and its
diagonalized form can be expressed [8] through fermionic occupation numbers:
\begin{equation}
H = \varepsilon_0  n_0 + \sum^{L - 1}_{k = 1}
\left( q + q^{-1} - 2 \cos \left( \frac{\pi k}{L} \right) \right) n_k \; ,
\end{equation}
where $\varepsilon_0 = 0$ is a zero fermionic mode and the
$n_i$'s $(i = 0, \cdots ,L-1)$ take the values $0$ and $1$. Barring accidental
degeneracies, each energy level of the Hamiltonian (2.6) is twice degenerate
due to the zero fermionic mode as long as $q$ is real. If $q$ is a root of
unity, additional degeneracies occur (see Ref. [8]).
The $V_i$'s defined by Eqs. (2.2),
(2.5) are the generators of the Hecke algebra:
\[
V_i V_{i \pm 1} V_i - V_i  =  V_{i \pm 1} V_i V_{i \pm 1} - V_{i \pm 1}
\]
\beqn
[ V_i , V_j ]  =  0 \; \; , \; \; | i-j | \geq 2
\eeqn
\[ V^2_i  =  (q + q^{-1}) V_i \qquad (i = 1, \ldots , L-1)
\]
corresponding to the quotient [9]:
\beqn
(V_i V_{i + 2}) V_{i + 1} (q + q^{-1} - V_i) (q + q^{-1} - V_{i +2}) = 0
\quad (i = 1, 2, \ldots , L - 3) \; .
\eeqn
{\it Case 1b}
\beqn
\Delta = \frac{q + q^{-1}}{2} \: , \;
h = 0 \: , \;
\alpha = \frac{q - q^{-1}}{2} \: , \;
\beta= -  \frac{q + q^{-1}}{4} \: .
\eeqn
The Hamiltonian is $U_q(sl(2))$-symmetric [10], the $V_i$'s are again the
generators of the Hecke algebra (2.7) but now correspond to the Temperley-Lieb
quotient:
\beqn
V_i V_{i \pm 1} V_i = V_i \; .
\eeqn
{\it Cases 2 and 3}
\beqn
(\Delta \pm h )^2 = 1 + \alpha^2 \; .
\eeqn
This symmetry is a bit unusual. It just says that the unique energy level with
$I_z = \frac{L}{2}$ has the same energy as one level with $I_z =
\frac{L}{2}-1$.
Alternatively the level with $I_z = - \frac{L}{2}$ has the same energy as one
level with $I_z = - \frac{L}{2} + 1$. For details see Appendix B.

At this point we will make a change of basis in the Hamiltonian (2.1) suggested
by the quantum chain formulation of dif\/fusion-reaction processes done in
Ref. [5]. For an $N$-state model ($N = 2$ for the Heisenberg chain, $N = 3$ for
the Hamiltonian (1.4)) we will denote the states by $m = 0, 1, \ldots , N-1$.
The state $m = 0$ will play a special role. In the space of $N \times N$
 matrices we
take the usual basis of matrices $E^{m n} (m, n = 0, 1, \ldots , N-1)$. The
matrix $E^{m n}$ has as only non-vanishing element the one in the
$m^{\hbox {th}}$ row
and $n^{\hbox {th}}$ column, and its value is 1. In this basis the Pauli matrix
 $\sigma^z$
can be written as
\beqn
\sigma^z = E^{00} - E^{11} = 1 - 2 E^{11} \; .
\eeqn
In the new basis the $V_i$'s of Eq. (2.2) have the following expression:
\begin{eqnarray}
V_i & = & - \left( E_i^{0 1} E^{1 0}_{i + 1} + E^{1 0}_i E^{0 1}_{i+1} \right)
+ v E^{1 1}_i E^{1 1}_{i + 1} \nonumber \\
& & + \: w \left( E^{1 1}_i + E^{1 1}_{i + 1} \right)
+ a \left( E^{1 1}_i - E^{1 1}_{i + 1} \right) + b \; ,
\end{eqnarray}
where
\beqn
v = - 2 \Delta \: , \;
w = \Delta + h \: , \;
a = \alpha \: , \;
b =  - \left( \frac{\Delta + \beta}{2} + h \right)  \; .
\eeqn

In order to construct an $N$-state system with the same spectrum, we write
non-diagonal terms that do not mix the states $m = 1, 2, \ldots , N-1$ and
diagonal terms where one permutes the states "$m$" (the state
"0" excluded). This takes us to the following Hamiltonian:
\beqn
H = \sum^{L - 1}_{i = 1} V_i \; ,
\eeqn
where
\[
V_i = - \sum^{N-1}_{m = 1} \left( E^{0 m}_{i} E^{m 0}_{i + 1} +
E^{m 0}_{i} E^{0 m}_{i + 1} \right)+ v \varepsilon^0_i \varepsilon^0_{i + 1} \]
\beqn
+ w \left( \varepsilon^0_i + \varepsilon^{0}_{i + 1} \right)
+ a \left( \varepsilon^0_i - \varepsilon^{0}_{i + 1} \right) + b
\eeqn
and
\beqn
\varepsilon^0 = \sum^{N-1}_{m = 1} E^{m m} \; .
\eeqn
In particular, if $N = 3$, we recover the $U_j$'s of Eq. (1.5). Since the
states "0" play a special role it is useful to introduce an operator
$Z$,
which counts them:
\beqn
Z = \sum^{L}_{i = 1} E^{0 0}_i = L  - \sum^L_{i = 1} \varepsilon^0_i \; ,
\eeqn
its eigenvalues $z$ are $0, 1, \ldots , L$. In particular if $N = 2$:
\beqn
Z = \frac{L}{2} + I^z \; ,
\eeqn
i.e., it coincides with the magnetization.

In Sec. 3 we will show, for $N \geq 2$, that the spectrum of the Hamiltonian
(2.15) is
independent of the number of states $N$. One corollary of this
observation (which has also been checked by direct calculation) is
that the $V_i$'s of Eq. (2.16),
if the
conditions (2.5) or (2.9) are satisfied,  also
satisfy the Hecke algebra (2.7) and the corresponding quotient relations
(2.8) or (2.10). We have to stress that all these results are valid for
open chains only. The chains with periodic boundary conditions have spectra
which depend on the number of states $N$ (see Sec. 3).

At this point it is useful to remind the reader of the phase diagram
[11] of the
Heisenberg chain, which is shown in Fig. 1. For $h = 0$, the system is massive
with a ferromagnetic ground state if $\Delta > 1$, massless and conformally
invariant if $- 1 \leq \Delta \leq 1$, and again massive with an
 antiferromagnetic
ground state if $\Delta < -1$. For a given $h$, the system is massive
ferromagnetic if $\Delta > 1-h$, then undergoes a Pokrovsky-Talapov [12]
transition at $\Delta = \Delta_{PT} = 1-h$, is in a massless incommensurate
 phase for
$\Delta_c < \Delta < 1 - h$, and reaches again the massive antiferromagnetic
phase at $\Delta < \Delta_c$. The $\Delta_c = \Delta_c (h)$ curve is
obtained by solving the equation
\beqn
h = 2 \sinh \theta \prod^\infty_{j = 1}
\left(\frac{1 - e^{j \theta}}{1 + e^{j \theta}}\right)^2 \; , \;
\Delta_c = - \cosh \theta \; .
\eeqn
Notice that the $\Delta = \Delta_{PT}$ line corresponds to the special symmetry
 related
to the condition (2.11) for $\alpha = 0$. The same phase diagram is valid
through the translation given by Eq. (2.14) for all the $N$-state systems
 described
by Eqs. (2.15) and (2.16), and therefore for the infinite repulsion limit
of the extended Hubbard model. In this case, in order to interpret the phase
diagram, it is better to work with a fixed value of the electron
 nearest-neighbour
interaction ($v$) and to vary by changing the fugacity ($w$). Phase (III)
 corresponds
to a phase where the fermions and holes are physically separated. In phase (I)
there are no fermions in the ground state, and in phase (II) we have fermions
and holes mixed.

\section{Energy levels and wave functions of the $N$-state model}
\setcounter{equation}{0}
In this section, we write the wave functions of the Hamiltonian (2.15), (2.16)
in terms of those of the Hamiltonian (2.1), (2.13), and deduce from this
that their spectra coincide. The degeneracies are also explained in terms
of the degeneracies of the $N = 2$ system (see also Sec. 6).

Since the Hamiltonian (2.1) commutes with the operator $Z$ (2.18), we can
choose its eigenstates among the eigenstates of $Z$, i.e.,  with definite
value for the number of states "0" (and hence for the number of states "1").
The Hamiltonian (2.15) also commutes with $Z$, and moreover it commutes
also with all the $ \sum_i  E_i^{m  m} $ for $ m = 1, \cdots , N - 1 $.
All the charges "$m$" are then conserved. It is actually easy to check that
the eigenstates of (2.1) are also eigenstates of (2.15). They define a
set of $2^L $ eigenstates of (2.15) that involve only the states "0" and "1".
So we already know that the spectrum of (2.1) is included in that of (2.15).

We will now introduce a set of operators that commute with the Hamiltonian
(2.15), and then  use them to construct all the eigenstates of (2.15)
by action on the previously defined eigenstates that involve only the states
"0" and "1".

Let us define
\beqn
J^{n m}_{(r,s)} \; ; \quad n , m  =  1, \cdots , N - 1 ; \qquad
r  =  1 , \cdots , L ; \quad
s  =  1, \cdots , r
\eeqn
as the operator with the following properties:
\begin{itemize}
\item[a)]
   $J^{n m }_{(r , s)} Z = Z J^{n m }_{(r , s)} \; = \;
( L - r ) J ^{n m }_{(r , s)}$,
i. e., $J^{n m }_{(r , s )}$ vanishes on states with a number of "0"
dif\/ferent from $L - r $.

\item[b)] $J^{n m }_{( r , s )}$ acts as $E^{n m }$ on the
$s^{\hbox{th}}$  site
on which the state is dif\/ferent from "0". (Note that $s \leq  r $ and that
we count from the left to the right.)
\end{itemize}
For instance,
\beqnar
J^{32}_{(4,3)} |201201 \rangle & = & |201301 \rangle \nonumber \\
J^{n m }_{(r , s)} |201201 \rangle & = & 0 \; \; {\mbox{if}} \; \; r \neq 4 \;
 .
\eeqnar
We then have
\beqn
\sum^L_{r = 1} \; \sum^r_{s = 1} \; \sum^{N - 1}_{m = 1}
\frac{1}{r}~J^{m m}_{(r , s)} \quad = \quad {\bf 1} - J_{(0)} \; ,
\eeqn
where $J_{(0)}$ is the projector on the state $ z = L$, i. e.,  with only
"0". By straightforward calculations one can show that the Hamiltonian
(2.15)
commutes with all the
$J^{n m }_{(r , s) } \quad (r = 1 , \cdots, L \; ; \; s = 1 , \cdots , r \; ;
\; n , m  = 1 , \cdots, N - 1 )$ and $J_{(0)}$. This property is only valid
for the open chain.

Consider now an eigenstate $\psi $ of the Hamiltonia (2.15),
involving only the states "0" and "1", and let $r$ be the eigenvalue of
$L - Z$ on it. Then the wave functions defined by
\beqn
\prod^{r}_{s = 1} J^{n_{s}, 1}_{(r , s)} \; \psi
\eeqn
are eigenstates of the Hamiltonian (2.15) with the same energy as $\psi$.
Here the $n_s$ take the set of values
$1, \cdots , N - 1 $ for $s = 1, \cdots , r$. The set of these wave functions
with $(n_1 , \cdots , n_r ) \in \{ 1 , \cdots , N - 1 \}^r$ is linearly
independent.

So each $\psi$ eigenstate of (2.1) leads to $(N - 1)^r$ eigenstates of (2.15).
The number of $\psi$'s such that $( L - Z ) \psi = r \psi $ being ${L \choose
 r}$,
we get a total of
\beqn
\sum\limits^{L}_{r = 0} {L \choose r} ( N - 1 )^r = N^L
\eeqn
eigenstates. Thus all the eigenstates are given by Eq. (3.4) and the spectrum
of (2.15) is identical with that of (2.1). This construction also provides
the degeneracies of eigenvalues of (2.15) in terms of those of (2.1): each
state of a multiplet of (2.1) provides $(N - 1 )^r $ states of the
corresponding multiplet of (2.15). We conclude this section with the following
observations:
\begin{itemize}
\item[a)] This construction does not work for periodic chains.
(The $J^{n m }_{(r , s)}$ do not commute with the Hamiltonian in this case.)
\item[b)] As explained in the following sections in the example of $N = 3$
the Hamiltonian (2.15) is in fact invariant under
$ U_q (\widehat{sl (N - 1)})$ for any $q$. To each site is
associated the sum of the trivial ($"0"$ state ) and the fundamental
($"1"$ to $"N - 1"$ states) representations. The previous construction can
be seen as a consequence of the fact that the iterated tensor products of the
fundamental representation of $U_q (\widehat{sl (N - 1)})$ are
irreducible [13] (See Appendix C).
\item[c)] The $N$-state generalization of the Hamiltonian with the constraints
(2.5) has the diagonal form given by Eq. (2.6) where now
$n_k (k = 1, \cdots , L - 1)$ takes the values one (one time) and zero
$(N - 1$ times). This is a generalization of the fermionic occupation number.
\end{itemize}

\section{Invariance of the Hamiltonian \hfil \break
under $U_q(sl(2))$ for all $q$}
\setcounter{equation}{0}
In the present and in the subsequent sections, we will discuss the
relationship between the spectra of the 2-state Hamiltonian (2.1), (2.13)
and the 3-state Hamiltonian (2.15), (2.16) from a dif\/ferent point of view,
namely, the $U_q ( \widehat{sl(2)})$-invariance of the latter.
As indicated, we choose $N = 3$, although our discussion holds in the
general case, too. As we first noticed by numerical diagonalizations,
the spectrum of the $3$-state Hamiltonian $H$ is highly degenerate. This
suggests that the symmetry of $H$ should be much higher than the obvious
$SU(2) \otimes U(1)$  invariance mentioned in the introduction. Visibly,
this invariance is generated at one site by the matrices
$\varepsilon^\pm , \; \varepsilon^z , \; \varepsilon^0$ specified in
 (1.7). Thus the
3-dimensional state space at one site carries the direct sum of a singlet
and a doublet representation of $SU(2)$ (corresponding to the indices $0$
and $1,2,$ respectively).

To gain some hints towards the symmetry of $H$ we sat about to construct a
$q$-analogue of this Hamiltonian. This is easily done, once one has noticed
that the doublets $(t^\pm )$ and $(t^{\prime \pm} )$ defined by
\beqn
t^+ \; = \; E^{10} \; , \qquad t^- \; = \; E^{20}
\eeqn
\beqn
t^{\prime +} \; = \; E^{02} \; , \qquad  t^{\prime -} \; = \; - E^{01}
\eeqn
are tensor operators of $SU (2)$ and that, up to the normalization,
\beqn
( t \otimes t^\prime )_0 \; = \; - E^{10} \otimes E^{01} \; - \; E^{20} \otimes
 E^{02}
\eeqn
\beqn
( t^\prime \otimes t )_0 \; = \;  E^{02} \otimes E^{20} \; + \; E^{01} \otimes
 E^{10}
\eeqn
are the scalar parts of the tensor product of $t$ and $t^\prime$ and of
$t^\prime $ and $t$, respectively. Notice that in Eq. (1.5) we put
$ \tau^- = t^+  \; ,\; \rho^- \; = \; t^- \; , \; \rho^+ \; = \; t^{\prime +}$
and $\tau^+ \; = \; -t^{\prime -}$.

To $q$-deform this structure we follow Ref.
[14]. We recall some of the definitions. Let $q$ be a non-zero complex number
with $q^2 \neq 1$. Then  $U_q (sl(2))$ is the universal algebra (associative,
with unit element) with generators $e , f , k^{\pm 1}$ and relations
\beqn
k k^{-1} \; = \; k^{-1} k = 1
\eeqn
\beqn
k e k^{-1} \; = \; q e \; , \qquad k f k^{-1}  = q^{-1} f
\eeqn
\beqn
e f - f e \; = \; \frac{k^2 - k^{-2}}{q - q^{-1}}
\eeqn
($e$ and $f$ correspond to the generators $\tilde{S}^\pm$ described in Sec. 1).
It is converted into a Hopf algebra by introducing the coproduct $\Delta$
through
\beqnar
\Delta (e) & = & e \otimes k^{-1} \;  + \;  k \otimes e \nonumber \\
\Delta (f) & = & f \otimes k^{-1} \;  + \;  k \otimes f  \\
\Delta ( k^{\pm 1} ) & = & k^{\pm 1} \otimes k^{\pm 1}, \nonumber
\eeqnar
the co-unit $\varepsilon$ through
\beqnar
\varepsilon ( e ) \; = \; \varepsilon ( f ) \; = \; 0 \nonumber \\
\varepsilon (k^{\pm 1} ) \; = \; 1 , \quad
\eeqnar
and the antipode $S$ through
\beqnar
S ( e ) \; &=& \; - q^{-1} e \nonumber \\
S ( f ) \; &=& \; - q f  \\
S ( k^{\pm 1} ) \; &=& \; k^{\mp 1} . \nonumber
\eeqnar
\vspace{0.5cm}
\\
It is well known that the spin $\frac{1}{2}$ representation of $U_q (sl(2))$
is undeformed (and, of course, so is the spin $0$ representation). Thus the
relevant representation of $U_q (sl(2))$ at one site is undeformed as well
and is given by
\beqn
e \rightarrow E^{12} \; , \quad
f \rightarrow E^{21} \; , \quad
k \rightarrow E^{00} \; + \; q^{1/2} E^{11} + q^{- 1/2} E^{22} .
\eeqn
Consequently, the deformation of the representation of $U_q (sl(2))$ in the
state space of $H$ is introduced by the non-trivial coproduct.

The $q$-deformed tensor operators $( t^\pm _q )$ and $(t^{\prime \pm }_q )$
are easily found to be
\beqn
t^+_q \; = \; t^+ \; = \; E^{10} \; , \qquad
t^-_q \; = \; t^-  \; = \; E^{20}
\eeqn
\beqn
t^{\prime +}_q \; = \; t^{\prime +} \; = \; E^{02} \; , \qquad
t^{\prime -}_q \; = \; q t^{\prime -}  \; = \; - q E^{01} .
\eeqn
Using the techniques of Ref. [14], we can then calculate the $q$-deformed
versions of Eqs. (4.3), (4.4). The surprising result is that with an adequate
normalization
\beqn
( t_q  \underline{\otimes} t^\prime_q )_0 \; = \; ( t \otimes t^\prime )_0
\eeqn
\beqn
( t^\prime_q  \underline{\otimes} t_q )_0 \; = \; ( t^\prime \otimes t )_0
\eeqn
(the underlining of $\otimes$ is to indicate that the tensor product is
constructed according to the rules for tensor operators over quasitriangular
Hopf algebras). Thus the scalar operators in Eqs. (4.3), (4.4) are undeformed.
This is not completely trivial since, again with a suitable normalization,
we find
\beqn
( t_q  \underline{\otimes} t_q )_0 \; = \; E^{10} \otimes E^{20} - q E^{20}
\otimes E^{10}
\eeqn
\beqn
( t^\prime_q  \underline{\otimes} t^\prime_q )_0 \; = \; E^{01} \otimes E^{02}
- q E^{02} \otimes E^{01} ;
\eeqn
thus, these scalar operators {\it are} deformed.

Since the remaining terms of $H$ are obviously $U_q (sl(2))$-invariant, we
conclude that the undeformed Hamiltonian $H$ is $U_q (sl(2))$-invariant,
for all $q$. In view of the well-known homomorphism of the $q$-deformed
af\/fine algebra $U_q (\widehat{sl(2)})$ onto $U_q (sl(2))$ (see the subsequent
section), it comes to the same to say that $H$ is even
$U_q (\widehat{sl(2)})$-invariant, for all $q$.

\section{Construction of $U_q (\widehat{sl(2)})$-invariant
\hfil \break Hamiltonians}
\setcounter{equation}{0}
The upshot of the foregoing section was that the Hamiltonian $H$ discussed
there is $U_q (\widehat{sl(2)})$-invariant. In the present section we are
going to rederive this result by means of a more systematic approach. We
shall start from a class of Hamiltonians generalizing $H$ and derive the
conditions under which they are $U_{q^\prime}
(\widehat{sl(2)})$-invariant for a
fixed value $q^\prime$ of $q$. This will lead us back to the Hamiltonian $H$.

To begin with, let us recall the definition of the Hopf algebra
$U_q (\widehat{sl(2)})$. With $q$ chosen as above, this is the universal
algebra (associative, with unit element) generated by
$e_i , f_i , k_i^{\pm 1} \; ; \; i \in \{ 1, 2 \} ,$ with relations
$( i , j \in \{ 1 , 2 \} )$:
\beqn
k_i k_i^{-1} = k_i^{-1} k_i = 1 \; \; , \qquad
k_i k_j = k_j k_i
\eeqn
\vspace{0.5cm}
\beqn
k_i e_j k_i^{-1} = q^{\frac{1}{2} a_{i j}} e_j \; \; , \qquad
k_i f_j k_i^{-1} = q^{- \frac{1}{2} a_{i j}} f_j
\eeqn
\vspace{0.5cm}
\beqn
e_i f_j - f_j e_i = \delta_{i j} \frac{k^2_i - k^{-2}_i}{q - q^{-1}}
\eeqn
\vspace{0.5cm}
\beqn
e^3_i e_j \; - \; [ 3 ]_q e^2_i e_j e_i \; + \; [ 3 ]_q  e_i e_j e^2_i \; - \;
e_j e^3_i \: = \; 0 \; , \; i \neq j
\eeqn
\vspace{0.5cm}
\beqn
f^3_i f_j \; - \; [ 3 ]_q f^2_i f_j f_i \; + \; [ 3 ]_q  f_i f_j f^2_i \; - \;
f_j f^3_i \: = \; 0 \; , \; i \neq j ,
\eeqn
where
\beqn
(a_{i j }) \; = \; {\; 2 \; -\!2 \choose -2 \quad 2}
\eeqn
is the Cartan matrix of the af\/fine Lie algebra $\widehat{sl (2)}$ and where,
for any integer $n$, we define
\beqn
[n]_q  =  \frac{q^n - q^{-n}}{q - q^{-1}} \quad .
\eeqn
The Hopf algebra structure is defined by introducing the coproduct $\Delta$
through
\beqnar
\Delta ( e_i ) & = & e_i \otimes k^{-1}_i \; + \; k_i \otimes e_i \nonumber \\
\Delta ( f_i ) & = & f_i \otimes k^{-1}_i \; + \; k_i \otimes f_i  \\
\Delta ( k^{\pm 1}_i ) & = & k^{\pm 1}_i \otimes k^{\pm 1}_i  \quad
,\nonumber
\eeqnar
the co-unit $\varepsilon$ through
\beqnar
\varepsilon ( e_i ) = \varepsilon ( f_i ) = 0 \nonumber \\
\varepsilon ( k^{\pm 1}_i ) = 1 \quad ,
\eeqnar
and the antipode $S$ through
\beqnar
S (e_i ) \; & = & - q^{-1} e_i \nonumber \\
S (f_i ) \; & = & - q f_i       \\
S (k_i^{\pm 1} ) & = & k^{\mp 1}_i  \; . \nonumber
\eeqnar
For our purposes it is important to recall that for any non-zero complex
number $c$ there exists a unique algebra homomorphism $g$ of
$U_q (\widehat{sl(2)})$ onto $U_q (sl(2))$ such that
\beqnar
g ( e_1 ) = e & , \qquad & g ( e_0 ) = c f  \nonumber \\
g ( f_1 ) = f & , \qquad & g ( f_0 ) = c^{-1} e     \\
g ( k_1 ) = k & , \qquad & g ( k_0 ) = k^{-1} \quad . \nonumber
\eeqnar
Consequently, if $\omega$ is any representation of $U_q (sl(2))$ in a vector
 space
$W$, then $\omega \circ g$ is a representation of $U_q (\widehat{sl(2)})$
in that vector
space. The coproduct of $U_q (\widehat{sl(2)})$ can then be used to construct
the $L^{\hbox{th}}$ tensorial power of $\omega \circ g$. We stress that this
 representation is
dif\/ferent from that  obtained by composing the $L^{\hbox{th}}$ tensorial
power of $\omega$ with $g$
[for $g$ is not a coalgebra homomorphism,
i.e., in general we have $\Delta g(a) \neq (g \otimes g) \Delta (a)$ for
$a \in U_q (\widehat{sl(2)})$ and hence, for example,
$(\omega \otimes \omega) \Delta g(a) \neq (\omega g \otimes \omega g)
\Delta (a)$].
Visibly, the
representation $\omega \circ g$ depends on $c$. In the following we
shall choose
$c = 1$.

After these preliminaries we proceed to our main topic. Let $A$ be any finite
index set not containing the numbers $1$ and $2$, and let $n$ be the number
of its elements. Later in this section we shall choose $A = \{ 0\}$. In the
following, the Greek indices $ \alpha , \beta , \gamma , \delta$ run through
$A \cup \{ 1 , 2\}$, and the Latin indices $a, b,  c, d$ through $A$.

We consider the 2-sites Hamiltonian
\beqn
H_{12} = \sum c_{\alpha \beta \gamma \delta} E^{\alpha \beta} \otimes
E^{\gamma \delta} ,
\eeqn
where the $c_{\alpha \beta \gamma \delta}$ are arbitrary complex numbers, the
$E^{\alpha \beta }$ are the basic \break
$(n + 2) \times (n + 2)$ matrices operating on
$W = {\bf C}^{n + 2}$ ({\bf C} denotes the field of complex numbers),
and where $H_{12}$ is an operator in $W \otimes W$.
We choose a non-zero complex number $q^\prime$ with $q^{\prime 2} \neq 1$
(later on we shall assume that $q^\prime$ is not a root of unity). According
to the foregoing discussion there exists a unique representation $\rho$ of
$U_{q^\prime} (\widehat{sl(2)})$ in $W$ such that
\beqnar
\rho (e_1 ) & = & \rho ( f_0 ) \; = \; E^{12} \nonumber \\
\rho (f_1 ) & = & \rho ( e_0 ) \; = \; E^{21}           \\
\rho (k_1 ) & = & \rho ( k_0^{-1} ) \; = \; K \; , \nonumber
\eeqnar
with
\beqn
K = \sum_a E^{a a} + q^{\prime \; 1/2} E^{11}  +
q^{\prime \; - 1/2} E^{22} .
\eeqn
Using the coproduct of $U_{q^\prime} (\widehat{sl(2)})$ we obtain the tensorial
square $(\rho \otimes \rho) \circ \Delta $ of $\rho $, i. e.,
the following representation of $U_{q^\prime} (\widehat{sl(2)})$ in
$W \otimes W$:
\beqnar
( \rho \otimes \rho ) \Delta (e_1 ) & = & E^{12} \otimes K^{-1} \; + \; K
 \otimes E^{12} \nonumber \\
( \rho \otimes \rho ) \Delta (f_0 ) & = & E^{12} \otimes K \; + \; K^{-1}
 \otimes E^{12} \nonumber \\
( \rho \otimes \rho ) \Delta (f_1 ) & = & E^{21} \otimes K^{-1} \; + \; K
 \otimes E^{21} \nonumber \\
( \rho \otimes \rho ) \Delta (e_0 ) & = & E^{21} \otimes K \; + \; K^{-1}
 \otimes E^{21}  \\
( \rho \otimes \rho ) \Delta (k_1 ) & = & K \otimes K \nonumber \\
( \rho \otimes \rho ) \Delta (k_0 ) & = & K^{-1} \otimes K^{-1} \; .
\nonumber
\eeqnar
One can show that $H_{12}$ commutes with $( \rho \otimes \rho ) \Delta (e_j )$
 and
$( \rho \otimes \rho ) \Delta (f_j )$; $j \in \{ 1, 2 \}$, if and only if
it takes the following special form:
\beqnar
H_{12} = &  & \sum_{a b c d} c_{a b c d} E^{a b} \otimes E^{c d} \nonumber \\
       & + & \sum_{a b }    c_{a b 1 1} E^{a b} \otimes (E^{11} + E^{22})
 \nonumber \\
       & + & \sum_{c d}     c_{1 1 c d} (E^{11} + E^{22}) \otimes E^{c d}
 \nonumber \\
       & + & \sum_{a d}     c_{a 1 1 d} (E^{a 1} \otimes E^{1 d} + E^{a 2}
                            \otimes E^{2 d}) \nonumber \\
       & + & \sum_{b c}     c_{1 b c 1} (E^{1 b} \otimes E^{c 1} + E^{2 b}
                            \otimes E^{c 2}) \nonumber \\
       & + &                c_{1 1 1 1} (E^{11} + E^{22}) \otimes
                            (E^{11} + E^{22}) \quad .
\eeqnar
We stress that the parameters appearing in this Hamiltonian are arbitrary
complex numbers. Thus, no $q^\prime$-dependence has been introduced by our
requirement for $U_{q^\prime} (\widehat{sl(2)})$-invariance. Consequently,
the Hamiltonian (5.16) is $U_q (\widehat{sl(2)})$-invariant for all
$q$.

It is easy to see that $H_{12}$ has the following more general symmetries.
Define
\beqn
D = q_0 \sum_a E^{aa} + q_1 E^{11} + q_2 E^{22}
\eeqn
\beqn
D^\prime = q_0 \sum_a E^{aa} + q^\prime_1 E^{11} + q^\prime_2 E^{22} ,
\eeqn
where $q_0 , q_1 , q^\prime_1 , q_2 , q^\prime_2$ are arbitrary complex
numbers. Then $H_{12}$ commutes with
\beqn
E^{12} \otimes D + D^\prime \otimes E^{12} \; , \quad
E^{21} \otimes D + D^\prime \otimes E^{21} \; , \quad D \otimes D ,
\eeqn
and, of course, also with $D^\prime \otimes D^\prime$. Obviously, this implies
that the corresponding $L$-sites Hamiltonian
\beqn
H = \sum^{L - 1}_{i = 1} H_{i , i + 1}
\eeqn
commutes with the corresponding symmetry operators in $W^{\otimes L}$, i.e.,
with
\beqnar
\sum_i {D^\prime} ^{\otimes (i-1)} \otimes E^{12} \otimes D^{\otimes (L-i)}  ,
 \nonumber \\
\sum_i {D^\prime} ^{\otimes (i-1)} \otimes E^{21} \otimes D^{\otimes (L-i)}
, \\
D^{\otimes L}
\; , \qquad  \quad
{D^\prime} ^{\otimes L}
\nonumber
\eeqnar
\vspace{0.5cm}
\\
The relevance will be illustrated in the following example.
We shall restrict our attention to the case $A = \{ 0 \}$;
furthermore, we shall assume that $c_{0110}$ and $c_{1001}$ are
dif\/ferent from
$0$. In that case, we may even assume that
\beqn
c_{0110} = c_{1001} = -1 ,
\eeqn
for up to the overall normalization of $H$ this can be achieved by a
similarity transformation in the state space $W^{\otimes L}$ (recall
that $W = {\bf C}^{n + 2} = {\bf C}^3)$, which preserves the special form
of $H$. With (5.22), $H$ takes the form (2.15), (2.16), with
\beqnar
v & = & c_{1111} - c_{1100} - c_{0011} + c_{0000} \nonumber \\
w & = & \frac{1}{2} (c_{1100} + c_{0011}) - c_{0000} \nonumber \\
a & = & \frac{1}{2} (c_{1100} - c_{0011})  \nonumber \\
b & = & c_{0000} \quad.
\eeqnar
Thus, choosing $A = \{0\}$ and assuming that (5.22) is satisfied, the
$U_{q^\prime} (\widehat{sl(2)})$-invariant Hamiltonians of the form (5.12)
are exactly the Hamiltonians introduced in (2.15), (2.16).

It is easy to see what the symmetry transformations (5.21) imply for the
spectral decomposition of $H$. Let $v_0 , v_1 , v_2$ denote the canonical
basis vectors of $W = {\bf C}^3$. For any subset
$J \subset \{ 1 , 2 , \cdots , L \}$, let $V_J$ denote the subspace of
\beqn
V = W^{\otimes L}\;,
\eeqn
which is spanned by the tensors
$v_{\alpha_1} \otimes \cdots \otimes v_{\alpha_L}$, with
$\alpha_i \in \{ 1 , 2 \}$ if $i \in J$, and $\alpha_i = 0$ otherwise.
Obviously, $V$ is the direct sum of the subspaces $V_J$ and
\beqn
\dim V_J = 2^r \quad {\mbox{with}} \quad r = \sharp J.
\eeqn
Furthermore, the $V_J$  are invariant under the transformations (5.21), and
it can be shown that these transformations act irreducibly on $V_J$.

The latter result can also be derived as follows. Visibly, $V_J$  carries the
$r^{\hbox{th}} $
tensorial power of the elementary spin $\frac{1}{2}$ representation
of $U_{q^\prime} (\widehat{sl(2)})$. But if $q^\prime$ is not a root of unity,
it is known [13] that this representation is irreducible.

The upshot is that $V = W^{\otimes L}$ decomposes into irreducible
$2^r$-dimensional  subspaces, with $r \in \{ 0 , 1, \cdots , L \}$,
and that exactly
$L \choose r$ of these subspaces have that dimension. This implies that the
eigenspaces of $H$ also decompose into irreducible subspaces of these
dimensions.

To be more precise, let $r \in \{ 0 , 1, \cdots , L \}$ and let
\beqn
V_r = \bigoplus\limits_{\sharp J = r} \; V_J \; .
\eeqn
Thus $V_r$ is the eigenspace of the operator $Z$ (defined in Eq. (2.18))
corresponding to the eigenvalue $L - r $. We stress that for $r$ fixed, the
irreducible representations acting in the $V_J$ with $\sharp J = r $ are
equivalent.

Furthermore, for $a \in \{ 1 , 2 \}$ we define
\beqn
V^a = ( {\bf C} v_0 + {\bf C} v_a)^{\otimes L}  \; ,
\eeqn
i.e., $V^a$ is spanned by the tensors of the form
$x_1 \otimes \cdots \otimes x_L$,
with $x_i \in \{ v_0 , v_a \} $ for all $i$. Finally, let
\beqn
V^a_r = V^a \cap V_r \; .
\eeqn
Obviously, we have
\beqn
V^a = \bigoplus\limits^{L}_{r = 0 }  V^a_r \; .
\eeqn
Furthermore, it is easy to see that $H$ maps the spaces $V^a$ and $V_r$
into themselves, hence also the spaces $V^a_r$. Thus, according to the
foregoing discussion, we only have to construct the spectral decomposition of
$H$ in the subspaces $V^a_r$ for one value of $a$, say $a = 1$. The action of
the symmetry transformations will then yield the corresponding decomposition
of $V_r$, hence of $V$. In particular, if the multiplicity of an eigenvalue of
$H$ in $V^1_r$ is equal to $m$, then the multiplicity of that eigenvalue
of $H$ in $V_r$ is equal to $m \cdot 2^r$.

Summarizing the preceding discussion we see that by means of the symmetry
transformations the spectral decomposition of $H$ in $V$ has been reduced
to that of $H$ in one of the subspaces $V^a$, i.e., to that of the 2-state
Hamiltonian (2.1), (2.13). This is in perfect agreement with the results of
Sec. 3. In particular, the existence of the symmetries $J^{nm}_{(r, s)}$
described in Sec. 3 follows directly from the results of the present
section. This will be explained in Appendix C.

\section{Multiplets of the $3^L$ chain when \newline
the underlying Heisenberg chain \newline
has supplementary symmetries}
\setcounter{equation}{0}
Consider the $3^L$ chain obtained with the construction of Sec. 5 from the
$2^L$ chain. We use the notation
\beqn
d_1 = w - a \; , \qquad d_2 = w + a \; , \qquad d_3 = v + 2w
\eeqn
for the diagonal parameters.
As explained in Appendix B, we have to consider four cases for the values of
$d_1 , d_2 , d_3$.
\\ \\
{\it Case 1a}
\beqn
d_1 = q^{-1} \quad , \qquad d_2 = q \quad , \qquad d_3 = q + q^{-1}\;.
\eeqn
The underlying $2^L$ chain then has the $U_{-q} (sl(1/1))$ symmetry, and its
states build doublets corresponding to adjacent eigenvalues
$r$ and $r - 1$ of $L - Z$. In the $3^L$ chain, the two vectors of each
doublet are highest-weight vectors of multiplets for the symmetry
$U_{q^\prime}(\widehat{sl(2)})$, with dimensions $2^r$ and $2^{r - 1}$. These
multiplets are gathered in a higher multiplet of dimension
$2^r + 2^{r - 1} = 3 \cdot 2^{r - 1}$. There are ${L - 1 \choose r - 1 }$
multiplets of this dimension. In the fermion picture, $L - 1 \choose r - 1$
is the number of states with $r - 1 $ fermions (among $L - 1$ fermions
with non-zero energy).
\\ \\
{\it Case 1b}
\beqn
d_1 = q^{-1} \quad , \qquad d_2 = q \quad , \qquad d_3 = 0\;.
\eeqn
The underlying $2^L$ chain has $U_q (sl(2))$ symmetry, and the
states are gathered into multiplets corresponding to the decomposition of
$( \hbox{Spin} \frac{1}{2} )^{\otimes L} $ with the fusion rules of
$U_q (sl(2))$. We shall consider only the case where $q$ is
not a root of unity.
[The case where it is  would easily follow from the decomposition
of $(\hbox{Spin} \frac{1}{2})^{\otimes L}$ into irreducible and indecomposable
representations.]
The multiplets of the $2^L$ chain collect states with different
eigenvalues of $L - Z$ as follows:
\beqnar
\begin{array}{ccccc}
\left\{ L , L - 1 , \cdots , 0 \right\} & , & [ L + 1 ] & , & ( 1 )
\nonumber \\
\left\{ L - 1 , L - 2 , \cdots , 1 \right\} & , & [ L - 1 ] & , & ( L - 1 )
\nonumber \\
& \vdots & \nonumber \\
\left\{ r , r - 1 , \cdots , L - r \right\} & , & [ 2r - L + 1 ] & , &
\left( {L \choose L - r} - {L \choose L - r - 1} \right)
\end{array}
\eeqnar
with $2r \geq L$. We have put in braces the eigenvalues of $L - Z$.
The brackets indicate the dimension of this $U_q (sl(2))$-multiplet,
whereas the parentheses indicates the number of multiplets of this type.

For the $3^L$ chain, each state of these $U_q (sl(2))$-multiplets
becomes a highest-weight vector of a $U_{q^\prime} (\widehat{sl
 (2)})$-multiplet.
We then have larger multiplets of dimensions
\beqnar
\begin{array}{ccccc}
2^L + 2^{L - 1} + \cdots + 2^0 & = & 2^{L + 1} - 1 & , & ( 1 ) \nonumber \\
2^{L - 1}+ 2^{L - 2} + \cdots + 2^1 & = & 2^L - 2 & , & ( L - 1 ) \nonumber \\
	& \vdots & \nonumber \\
2^r + 2^{r - 1} + \cdots + 2^{L - r} & = & 2^{r + 1} -2^{L - r} & , &
\left( {L \choose L - r} - { L \choose L - r - 1} \right)
\end{array}
\eeqnar
where again, in the parenthesis, we indicate the numbers of these multiplets.
\\ \\
{\it Case 2}
\beqn
d_1 d_2 = 1 \quad , \quad (d_1 - d_3 ) (d_2 - d_3) \neq 1 \;.
\eeqn
As shown in Appendix B, the supplementary symmetry builds a unique
2-dimensional
 multiplet
of the $2^L$ chain with the state $v_0^{\otimes L}$
[for the notation, see the lines preceding Eq. (5.24)] and a state
with the
eigenvalue of $Z$ equal to $L - 1$. In the $3^L$ chain, these two states
become highest-weight vectors of $U_{q^\prime} (\widehat{sl(2)})$-multiplets
of dimension 1 and 2, respectively. So we finally get a unique multiplet
of dimension three, while the other
$U_{q^\prime} (\widehat{sl(2)})$-multiplets are not combined.
\\ \\
{\it Case 3}
\beqn
d_1 d_2 \neq 1 \quad , \quad (d_1 -d_3 ) \; (d_2 -d_3 ) = 1 \;.
\eeqn
The two states that build a doublet in the $2^L$ chain have now
$Z$-eigenvalues $0$ and $1$. In the $3^L$ chain problem, these states are
highest-weight vectors of $U_{q^\prime} (\widehat{sl(2)})$-multiplets of
 dimensions
$2^L$ and $2^{L-1}$. We then have a multiplet of dimension $3 \cdot 2^{L - 1}$.
All the other states remain in multiplets of dimension $2^r$ $(0 \leq r < L)$.
\section{Conclusions}
We have started our paper with a physically interesting one-dimensional system
of fermions with an interaction given by three coupling constants
$\frac{U}{2t}, \frac{V}{2t}$ and $\frac{W}{2t}$. The phase diagram of this
 system is unknown.
One rigorous result of our investigation is that in the strong repulsion limit,
the phase diagram coincides with the one of the Heisenberg chain in a magnetic
field (see Fig. 1). The Eq. (2.14) of Sec. 2 describes this mapping. We also
give an interpretation of the phase diagram of Fig. 1, which is usually
understood in the language of magnetic systems, in terms of hard-fermion
physics (see Appendix A). Our work has gone beyond this investigation.
On the one hand in Sec. 2 we
have obtained $N$-state models whose spectra coincide with those of the
Heisenberg model (the $N = 3$ case being relevant to the fermionic problem).
The eigenfunctions of these Hamiltonians can be obtained from those of the
$N = 2$ case using the construction given in Sec. 3 (see also Appendix C). On
 the
other hand we have noticed (see Sec. 4) that for $N = 3$ the multiplet
structure
of the chains is related to finite-dimensional representations of the af\/fine
algebra $U_{q}(\widehat{sl(2)})$, the chains being invariant under this
algebra for any $q$. In Sec. 5 we give the conditions under which an
($n+2$)-state
Hamiltonian is invariant under $U_q(\widehat{sl(2)})$ for any $q$ and find a
new set of Hamiltonians whose spectra are no longer given by the
Heisenberg model. Finally, we would like to notice that for the Heisenberg
model we have found a new symmetry valid for finite chains when the
parameters verify the relations (2.11). This symmetry is discussed in Appendix
 B. In
particular, this symmetry applies to the Pokrovsky-Talapov line.

\appendix
\renewcommand{\theequation}{\Alph{section}.\arabic{equation}}
\renewcommand{\thesection}{\mbox{Appendix }\Alph{section}}
\setcounter{equation}{0}

\section{From 4-state fermionic to \newline
3-state magnetic spin chains}

In this appendix we obtain the Hamiltonian (1.3)--(1.5)
as the $U \to \infty$
limit of the fermionic Hamiltonian (1.1), (1.2).
Since in this limit double
occupancy of fermions is forbidden (it costs an infinite energy to put
fermions at the same site) we can start our derivation by constructing a
Fock space where, in a given lattice point, we may have no fermions ("0"),
a fermion with up spin ("$\uparrow$") or down spin ("$\downarrow$").
Due to the fermionic nature of the operators appearing in (1.3)--(1.5)
we choose
an order and construct directly a representation in the $3^L$-dimensional
Fock space. The chosen order is illustrated in the following two examples
\begin{samepage}
\begin{eqnarray}
| \uparrow , \uparrow , \downarrow , 0 , \uparrow , \ldots , \uparrow \rangle
& = &
c^+_{1, \uparrow} c^+_{2, \uparrow} c^+_{3, \downarrow} c^+_{5, \uparrow}
\ldots c^+_{L, \uparrow} |0, 0, 0, \ldots , 0 \rangle \nonumber \\*[.5cm]
| 0, \uparrow ,  \downarrow , 0 , \uparrow , \ldots  \rangle
& = &
c^+_{2, \uparrow} c^+_{3, \downarrow} c^+_{5, \uparrow}
\ldots  |0, 0, 0, \ldots , 0 \rangle \; ,
\end{eqnarray}
where
\beqn
c_{i, \sigma} | 0, 0, 0, \ldots \rangle = 0 \: ,
\qquad i = 1, 2, \ldots , L \; ; \; \sigma = \uparrow , \downarrow \; .
\eeqn
With this convention we can represent the Hamiltonian (1.1), (1.2) as \\
\arraycolsep0.3mm
\parbox{7cm}{\begin{eqnarray*}
H^\prime_0 = \sum^{L-1}_{i = 1} {\bf 1}_1 \otimes \ldots \otimes {\bf 1}_{i-1}
 \otimes
&
\left[
\begin{array}{*{11}{c}}
d_0 &      &     & \vdots &    &   &   & \vdots &   &   &  \\
    & d_1  &     & \vdots & -1 &   &   & \vdots &   &   &  \\
    &      & d_1 & \vdots &    &   &   & \vdots & -1&   & \\
\multicolumn{11}{c}\dotfill                      \\
    & -1   &     & \vdots & d_2 &     &     & \vdots &   &   &  \\
    &      &     & \vdots &     & d_3 &     & \vdots &   &   &  \\
    &      &     & \vdots &     &     & d_3 & \vdots &   &   &  \\
\multicolumn{11}{c}\dotfill                      \\
    &      &  -1 & \vdots &     &     &     & \vdots & d_2  &   &  \\
    &      &     & \vdots &     &     &     & \vdots &   & d_3  &  \\
    &      &     & \vdots &     &     &     & \vdots &   &   & d_3
\end{array}
\right]_{i, i+1}
\!\!\!\!\!\!\!\otimes \ldots \otimes {\bf 1}_L
\end{eqnarray*}} \hfill
\parbox{1cm}{\begin{eqnarray}\end{eqnarray}}
\end{samepage}
\newpage
\arraycolsep0.3mm
\parbox{7cm}{\begin{eqnarray*}
H^\prime_1 = \frac{G}{2t} \sum^{L-1}_{i = 1} {\bf 1}_1 \otimes \ldots \otimes
 {\bf 1}_{i-1} \otimes
&
\left[
\begin{array}{*{11}{c}}
0 &    &     & \vdots &    &   &   & \vdots &   &   &  \\
  & 1  &     & \vdots &    &   &   & \vdots &   &   &  \\
  &    & -1  & \vdots &    &   &   & \vdots &   &   & \\
\multicolumn{11}{c}\dotfill                      \\
  &    &     & \vdots & 1  &     &     & \vdots &   &   &  \\
  &    &     & \vdots &    &  2  &     & \vdots &   &   &  \\
  &    &     & \vdots &    &     & 0   & \vdots &   &   &  \\
\multicolumn{11}{c}\dotfill                      \\
  &    &     & \vdots &    &     &     & \vdots & -1  &   &  \\
  &    &     & \vdots &    &     &     & \vdots &     & 0 &  \\
  &    &     & \vdots &    &     &     & \vdots &     &   & -2
\end{array}
\right]_{i, i+1}
\!\!\!\!\!\!\!\otimes \ldots \otimes {\bf 1}_L \; ,
\end{eqnarray*}} \hfill
\parbox{1cm}{\begin{eqnarray}\end{eqnarray}} \\
where
$d_0 = 0 , d_1 = \frac{W}{2t} -  \frac{A}{2t} \: , \:
 d_2 = \frac{W}{2t} + \frac{A}{2t} \: , \:
 d_3 = \frac{V}{t} + \frac{W}{t} $ and ${\bf 1}_i$
is a $3 \times 3$ unit matrix attached at site $i$. The elements not shown
explicitly in the above matrices have zero value.

This Hamiltonian can promptly be rewritten in terms of the matrices
introduced in (1.7) \\
\begin{eqnarray}
H^\prime_0  = && \sum^{L-1}_{i = 1} {\mbox{\Huge\{}}
- \left( \varrho^-_i \varrho^+_{i + 1} + \varrho^+_i \varrho^-_{i + 1}
+        \tau^+_i \tau^-_{i + 1} + \tau^-_i \tau^+_{i + 1} \right)
\nonumber
\\*[0.2cm]
& + &  \left.
 \frac{V}{t} \varepsilon^0_i \varepsilon^0_{i + 1}
+ \frac{W}{2t} \left( \varepsilon^0_i  + \varepsilon^0_{i + 1} \right)
+ \frac{A}{2t} \left( \varepsilon^0_i  - \varepsilon^0_{i + 1} \right) + b
\right\}
\end{eqnarray}
\begin{equation}
H^\prime_1 = - \frac{G}{2t} \sum^L_{i = 1} \varepsilon^z_i \; .
\end{equation}
The reader has already noticed that our proof of the connection between the
Hamiltonians (1.3)--(1.5) and (1.1) and (1.2) applies to open chains only. In
the case of a periodic chain one has to split the sectors of the fermionic
Fock space according to the parity of the fermionic number. Sectors with
an odd (even) number of fermions will correspond to the Hamiltonian
(1.3)--(1.5)
with periodic (antiperiodic) boundary condition.

\section{New symmetries of \newline
         the Heisenberg model}
\setcounter{equation}{0}
Let us consider the $2^L$ state Hamiltonian $H = \sum^{L-1}_{j = 1} V_j$
with $V_j$ given by Eq. (2.13); $V_j$ can also be written
\begin{eqnarray}
V_j = b \: + &
\left(
\begin{array}{*{7}{c}}
0 & {} &   & \vdots &        & {} & \\
  & {} & d_1 & \vdots  & -1  & {} & \\
\multicolumn{7}{c}\dotfill     \\
  & {}  &  -1 & \vdots & d_2 & {} & \\
  & {}  &     & \vdots &   &   {} &d_3
\end{array} \right)
\end{eqnarray}
with $d_1 = w - a \: , \: d_2 = w + a \: , \: d_3 = v + 2w$.

We look for symmetries of $H$ of the form
\beqn
\tilde{e} = \sum^{L}_{j = 1} D^{\prime \otimes (j - 1)}
\otimes E^{0 1} \otimes D^{\otimes (L-j)} \; ,
\eeqn
where $D$ and $D^\prime$ are diagonal matrices and
$D^{\otimes (j-1)} = D \otimes D \otimes \ldots \otimes D$ \break
($j-1$ times).

If $H$ commutes with $\tilde{e}$ given by (B.2), it will also commute with
$\tilde{f}$, its transpose, given by
\beqn
\tilde{f} = \sum^L_{j = 1} D^{\prime \otimes (j-1)} \otimes E^{1 0}
\otimes D^{\otimes (L-j)} \; .
\eeqn
The existence of these symmetries depends on the values of $d_1 , d_2 , d_3$.
Let us write
\begin{eqnarray}
D =
\left(
\begin{array}{*{3}{c}}
l_{0} & {} & 0 \\
0   & {} & l_{1}
\end{array}
\right)\; , \qquad
D^\prime =
\left(
\begin{array}{*{3}{c}}
l^\prime_{0} & {} & 0 \\
0          & {} & l^\prime_{1}
\end{array}
\right).
\end{eqnarray}
Then the conditions on $l_i, l^\prime_i , d_1 , d_2 , d_3$ for
$[H , \tilde{e} ] = 0$ are \\
\parbox{11cm}{\begin{eqnarray*}
d_1  l^\prime_0 - l_0 = 0 & \quad , \quad & l^\prime_0 - d_2 l_0 = 0 \\
(d_1 - d_3) l_1 - l^\prime_1 = 0 & \quad , \quad & l_1 - (d_2 - d_3) l^\prime_1
 = 0 \; .
\end{eqnarray*}} \hfill
\parbox{1cm}{\begin{eqnarray}\end{eqnarray}} \\
The solvability of these equations depends on the values of
$d_1 , d_2 , d_3$, and we have to consider four cases: \\ \\
{\it Case 1}
\beqn
d_1 d_2 = 1 \quad , \qquad (d_1 - d_3) (d_2 - d_3) = 1
\eeqn
i.e., either
$$
d_1 = q^{-1} \quad , \qquad d_2 = q \quad , \qquad d_3 = q + q^{-1}
\eqno({\hbox {B.7a}})
$$
or
$$
d_1 = q^{-1} \quad , \qquad d_2 = q \quad , \qquad d_3 = 0 \; .
\eqno({\hbox {B.7b}})
$$
These cases correspond to the conditions (2.5) and (2.9), respectively,
and the symmetries generated by $\tilde{e}$ and $\tilde{f}$ can be
identified with $U_{-q}(sl(1/1))$ and $U_q(sl(2))$, respectively. \\ \\
{\it Case 2}
\setcounter{equation}{7}
\beqn
d_1 d_2 = 1 \quad , \qquad (d_1 - d_3)(d_2 - d_3) \neq 1 \;.
\eeqn
Then
\begin{eqnarray}
D =
\left(
\begin{array}{*{3}{c}}
l_0 & {} & 0 \\
0   & {} & 0
\end{array}
\right)\; , \qquad
D^\prime =
\left(
\begin{array}{*{3}{c}}
d_2 l_0 & {} & 0 \\
0        & {} & 0
\end{array}
\right).
\end{eqnarray}
Dropping an overall factor $l^{L-1}_0$, we have
\beqn
\tilde{e} = \sum^L_{j = 1}  d_2^{j-1} (E^{00})^{\otimes (j-1)}
\otimes E^{01} \otimes (E^{00})^{\otimes (L-j)}
\eeqn
\beqn
\tilde{f} = \sum^L_{j = 1} d_2^{j-1} (E^{00})^{\otimes (j-1)}
\otimes E^{10} \otimes (E^{00})^{\otimes (L-j)}  \; .
\eeqn
We then see that all the states are cancelled by $\tilde{f}$, except the state
 with
charges "$0$" only. In fact, $\tilde{e}$ and $\tilde{f}$ connect the states
$| 0 \ldots 0 \rangle$ and $\tilde{f} | 0 \ldots 0 \rangle$ in a doublet, and
all the remaining states are singlets.

Two levels are then degenerate, the others being non-degenerate. This
degeneracy
of two levels only has also been checked numerically. \\ \\
{\it Case 3}
\beqn
d_1 d_2 \neq 1 \quad , \qquad (d_1 - d_3)(d_2 -  d_3) = 1
\eeqn
The situation is similar to the previous case and we can thus write
\beqnar
\tilde{e} = \sum^L_{j = 1} \left( d_1 - d_3 \right)^{j - 1} \left( E^{11}
 \right)^{\otimes (j - 1)}
\otimes E^{01} \otimes \left( E^{11} \right)^{\otimes (L - j)} \\
\tilde{f} = \sum^L_{j = 1} \left( d_1 - d_3 \right)^{j - 1} \left( E^{11}
 \right)^{\otimes (j - 1)}
\otimes E^{10} \otimes \left( E^{11} \right)^{\otimes (L - j)} .
\eeqnar
Now $\tilde{e}$ vanishes on all the states but $| 1 \ldots 1 \rangle$, and
$\left\{ |1 \ldots 1 \rangle , \tilde{e} | 1 \ldots 1 \rangle \right\}$ build
a doublet. All the other states are singlets for $\tilde{e}, \tilde{f}$.
Again, only two levels coincide. \\ \\
{\it Case 4}
\beqn
d_1 d_2 \neq 1 \quad , \qquad (d_1 - d_3) (d_2 - d_3) \neq 1 \; .
\eeqn
Then $\tilde{e} = \tilde{f} = 0$; we get no further symmetry.
\\ \\ \\
The symmetries that exist for the $2^L$ state Hamiltonian in the cases
(1), (2), and (3) are combined with the other symmetries
[$U_{q^\prime}\widehat{(sl(N)})$ in the $(N + 1)^L$ state case].
In particular, when two states build a doublet of the $2^L$ state Hamiltonian,
the representations of $U_{q^\prime}\widehat{(sl(N)})$ associated with these
 states in the
$(N + 1)^L$-state case are also combined into higher multiplets.
In the case (2), we get a $1 + N$-dimensional multiplet, together with
$N^r$-dimensional multiplets.
In the case (3), we get a $N^L + N^{L-1}$-dimensional multiplet, together
with $N^r$-dimensional multiplets (see Sec. 6).

\section{The symmetries $J^{mn}_{(r, s)}$ \newline reconsidered}
\setcounter{equation}{0}
In the present appendix we want to show that the existence of the symmetries
$J^{mn}_{(r,s)}$ described in Sec. 3 follows from the results of Sec. 5
by using the standard representation theory of associative algebras. In doing
so we use the notation introduced in Sec. 5.

Let $\rho^L$ be the $L^{\hbox{th}}$ tensorial power of the
representation $\rho$ of
$U_q(\widehat{sl(2)})$ as specified in Eq. (5.13) (we simplify the notation and
write $q$ instead of $q^\prime$). Thus $\rho^L$ acts in the state space
$V = W^{\otimes L}$ of the $L$-sites chain, and we know that $V$ decomposes
into the direct sum of the invariant irreducible subspaces
$V_J , J \subset \{1, 2, \ldots , L \} $. Obviously, if
$J, J^\prime \subset \{ 1, 2, \ldots , L\}$
and if $\sharp J$ and $\sharp J^\prime$ are different, the
representations
given by $V_J$ and $V_{J^\prime}$ are inequivalent. On the other hand, if
$\sharp J = \sharp J^\prime = r$, these representations are equivalent. More
precisely, if we forget about the tensorial factors $v_0$ (which can be done
by use of canonical $U_q\widehat{(sl(2))}$-module isomorphisms) they simply
coincide with the $r^{\hbox{th}}$ tensorial power of the elementary
2-dimensional
representation of $U_q\widehat{(sl(2))}$, acting in
$({\bf C} v_1 + {\bf C} v_2)^{\otimes r}$.

Now the representation theory of associative algebras tells us the following.
Choose $r \in \{ 0, 1, \ldots , L \} $ and let $A$ be an arbitrary linear
 operator
in $({\bf C} v_1 + {\bf C} v_2)^{\otimes r}$. Then there exists an element
$a \in U_q(\widehat{sl(2)})$ such that $\rho^L (a)$ acts on $V_J$ as $A$ if
$\sharp J = r$, but as $0$ if $\sharp J \neq r$.

Of course, all these operators $\rho^L (a)$ commute with the Hamiltonian
$H$; moreover, it is easy to construct the operator $\rho^L (a)$
explicitly. Without loss of generality, we may assume that
\beqn
A = A_1 \otimes \ldots \otimes A_r \; ,
\eeqn
where $A_1 , \ldots , A_r$ are arbitrary $2 \times 2$ matrices, with matrix
indices in $ \{ 1,2 \} $. Let $\bar{A}_s$ be the $3 \times 3$ matrix obtained
 from
$A_s$ by adding a zeroth row and column consisting of zeros only. Define the
operator $J_r (A_1 , \ldots , A_r)$, acting in $V$, as follows:
\beqn
J_r (A_1 , \ldots , A_r ) = \sum E^{00} \otimes \ldots \otimes \bar{A}_1
\otimes \ldots \otimes \bar{A}_2 \otimes \ldots \otimes \bar{A}_r  \otimes
\ldots \otimes E^{00}  \; ,
\eeqn
where the sum is extended over all tensor products of the matrices
$\bar{A}_1 , \ldots , \bar{A}_r$ and $L-r$ matrices $E^{00}$, with the
$ \bar{A}_1 , \ldots , \bar{A}_r$ appearing in their given order. Then
$J_r (A_1 , \ldots , A_r)$  is the operator acting on
$V_J$ as $A_1 \otimes \ldots \otimes A_r$ if $\sharp J = r$ and as $0$ if
$\sharp J \neq r$. Thus we know that the operator $J_r (A_1 , \ldots , A_r)$
commutes with the Hamiltonian $H$.

We note that if $r^\prime \in \{ 0 , 1, \ldots , L \} $ and if
$A^\prime_1 , \ldots , A^\prime_{r^\prime}$ is a second family of $2 \times 2$
matrices, we have
\begin{equation}
\begin{array}{rcll}
J_r (A_1 , \ldots , A_r) J_{r^\prime} (A^\prime_1 , \ldots ,
 A^\prime_{r^\prime})
& = &
J_r  (A_1 A^\prime_1 , \ldots , A_r A^\prime_r) & \quad  {\mbox{if}} \; \;
r = r^\prime
\\
& = &
0 & \quad  {\mbox{if}} \; \; r \neq r^\prime \; .
\end{array}
\end{equation}
Moreover, our results show that the operator algebra
$\rho^L (U_q(\widehat{sl(2)}))$ coincides with the linear span of the set of
all operators $J_r (A_1 , \ldots , A_r)$. In particular, the operators
$\rho^L (e_i ) , \rho^L (f_i ) , \rho^L (k^{\pm 1}_i) ; i \in \{1 , 2 \}$,
can be written in this form. For example, by definition of $\rho^L$, we
have
\beqn
\rho^L (e_1 ) = \sum^L_{r = 1} \: \sum^r_{s = 1} \:
J_r(k, \ldots , k , E^{12} , k^{-1} , \ldots , k^{-1} ) \; ,
\eeqn
where $E^{12}$ is in position $s$ and $k$ denotes the $2 \times 2$ matrix
\beqn
k = q^{1/2} E^{11} + q^{-1/2} E^{22} \; .
\eeqn
On the other hand, it is easy to see that
\beqn
J^{mn}_{(r, s)} = J_r (1, \ldots , E^{mn} , \ldots , 1) \; ,
\eeqn
where, now, $E^{mn}$ is in position $s$ and $1$ denotes the $2 \times 2$ unit
matrix. In view of Eq. (C.3), this implies that the operators $J^{mn}_{(r,s)}$
and $J_{(0)}$ generate the associative operator algebra
$\rho^L (U_q(\widehat{sl(2)}))$.

\end{document}